\documentclass[aps,pre,preprint,groupedaddress,showpacs]{revtex4}
\usepackage{graphicx}

\begin{document}

\title{Formation of a Standing-Light Pulse through Collision of Gap Solitons}
\author{William C. K. Mak$^{1}$, Boris A. Malomed$^{2,1}$, and Pak L. Chu$^1$}

\affiliation{$^1$Optoelectronic Research Centre, Department of Electronic
Engineering, City University of Hong Kong}

\affiliation{$^2$Department of Interdisciplinary Studies,
Faculty of Engineering, Tel Aviv University,
Tel Aviv 69978, Israel}

\begin{abstract}
Results of a systematic theoretical study of collisions between moving
solitons in a fiber grating are presented. Various outcomes of the collision
are identified, the most interesting one being merger of the solitons into a
single zero-velocity pulse, which suggests a way to create pulses of
``standing light''. The merger occurs with the solitons whose energy takes
values between $0.15$ and $0.35$ of the limit value, while their velocity is
limited by $\approx 0.2$ of the limit light velocity in the fiber. If the
energy is larger, another noteworthy outcome is acceleration of the solitons
as a result of the collision. In the case of mutual passage of the solitons,
inelasticity of the collision is quantified by the energy-loss share. Past
the soliton's stability limit, the collision results in strong deformation
and subsequent destruction of the solitons. Simulations of multiple
collisions of two solitons in a fiber-loop configuration are performed too.
In this case, the maximum velocity admitting the merger increases to
$\approx 0.4$ of the limit velocity. Influence of an attractive local defect
on the collision is also studied, with a conclusion that the defect does not
alter the overall picture, although it traps a small-amplitude pulse.
Related effects in single-soliton dynamics are considered too, such as
transformation of an input sech signal into a gap soliton (which is
quantified by the share of lost energy), and the rate of decay of a
quiescent gap soliton in a finite fiber grating, due to energy leakage
through loose edges.
\end{abstract}


\maketitle

\section{Introduction}

Bragg gratings (BGs) are structures in the form of a periodic variation of
the core refractive index, which are written on a fiber or other optical
waveguide \cite{Kashyap}. Devices based on fiber gratings, such as filters
and gain equalizers, are among the most widely used components of optical
systems. \textit{Gap solitons} (in a more general context, they are called
BG solitons \cite{Sterke}) exist in fiber gratings due to the interplay
between the BG-induced effective dispersion and Kerr nonlinearity of the
fiber material. Exact analytical solution for BG solitons in a standard
model were found in Refs. \cite{Aceves,Demetri}, and their stability was
studied later, showing that, approximately, half of them are stable
(see details below) \cite{Tasgal,stability}.
Spatial solitons and their stability in a model of planar
BG-equipped waveguide, taking into regard two polarizations of light, were
recently considered in Ref. \cite{Skryabin}.

Lately, a lot of attention has been attracted to possibilities of capturing
``slow light'' \cite{slowlight}, and, in particular, of slowly moving
optical solitons \cite{slowsoliton} in various settings. Fiber gratings are
natural candidates for a nonlinear medium where it is potentially possible
to stop the light, as zero-velocities BG solitons, in which the left- and
right-traveling waves are in permanent dynamical equilibrium, are available
as exact solutions \cite{Aceves,Demetri}, and a part of them are stable \cite
{Tasgal,stability}. Actually, BG solitons that were thus far observed in the
experiment were fast ones, moving at a velocity $\approx 75\%$ of the limit
light velocity in the fiber \cite{experiment}. A possible way to create a
zero-velocity soliton is to use an attractive finite-size \cite{Weinstein}
or $\delta $-like \cite{we} local defect in the BG which attracts solitons
(it was demonstrated in Ref. \cite{trapping} that a defect can also
stimulate a nonlinear four-wave interaction without formation of a soliton).
Moreover, it is possible to combine the attractive defect with local gain,
which opens a way to create a permanently existing pinned soliton, even in
the presence of loss \cite{we2}.

One of objectives of this paper is to explore a possibility of slowing down
BG solitons by colliding two identical ones moving in opposite directions in
the fiber grating. Collisions are quite feasible from the experimental
standpoint, as a characteristic length necessary for the formation of a BG
soliton is $\simeq 2$ cm \cite{experiment}, while uniform fiber gratings
with a length $1$ m or even longer are now available. Already in the first
work \cite{Aceves}, where exact solutions for the moving solitons were
found, their collisions were simulated, with a conclusion that they passed
through each other, re-appearing with intrinsic vibrations, which may be
explained by excitation of an intrinsic mode which a stable BG soliton
supports \cite{Tasgal}. Note that broad small-amplitude BG solitons are
asymptotically equivalent to nonlinear-Schr\"{o}dinger (NLS) solitons, hence
collisions between them are completely elastic \cite{Litchinitser}. However,
in a more generic case results may be different, as the standard
fiber-grating model, see Eqs.~(\ref{pdes}) below, is not an integrable one,
on the contrary to the NLS equation. Systematic simulations are thus needed
to study head-on collisions between BG solitons, results of which are
reported below in Section III, after presenting the model in Section II. The
main finding is that, at relatively small values of the solitons's
velocities $\pm c$, and not too large values of the solitons' energy, the
solitons merge into a single \emph{standing} one. In the case when the
solitons pass through each other, we quantify the collision by an
energy-loss share. In section IV, we report results of simulations of
multiple collisions between two solitons, to model a situation in a fiber
loop. These results show that multiple collisions essentially increase the
maximum velocity which admits the merger. Simulations were also carried out
to check if inclusion of a local defect attracting the solitons may assist
the fusion of the colliding solitons. In Section V we demonstrate that the
defect does not affect the situation essentially; however, a small-amplitude
trapped pulse, which captures a relatively small share of the initial
solitons' energy, appears as a result of the collision. Finally, in Section
VI we report some related results pertaining to single-soliton dynamics,
viz., reshaping of an input pulse of a sech form (as suggested by the NLS
equation) into a BG soliton in the fiber grating, and gradual decay of a
soliton in a finite-length grating due to the energy leakage through open
ends. Section VII concludes the paper.

\section{The model}

The commonly adopted model of nonlinear fiber gratings is based on a system
of coupled equations for the right- ($u$) and left- ($v$) traveling waves
\cite{Sterke},

\begin{eqnarray}
iu_{t}+iu_{x}+v+\left[ \left( 1/2\right) |u|^{2}+|v|^{2}\right] u &=&0,
\nonumber \\
iv_{t}-iv_{x}+u+\left[ \left( 1/2\right) |v|^{2}+|u|^{2}\right] v &=&0,
\label{pdes}
\end{eqnarray}
where $x$ and $t$ are the coordinate and time, which are scaled so that the
linear group velocity of light is $1$, the Bragg-reflectivity coefficient
being $1$ too. Exact solutions to Eqs. (\ref{pdes}), which describe solitons
moving at a velocity $c$ ($c^{2}<1$), were found in Refs. \cite{Aceves} and
\cite{Demetri}:

\begin{eqnarray}
u &=&\alpha W(X)\mathrm{exp}\,\left[ y/2+i\phi (X)-iT\cos \theta +i\phi _{0}
\right] ,  \nonumber \\
v &=&-\alpha W^{\ast }(X)\mathrm{exp}\,\left[ -y/2+i\phi (X)-iT\cos \theta
+i\phi _{0}\right] .  \label{movsol}
\end{eqnarray}
\noindent Here, $\theta $ is an intrinsic parameter of the soliton family
(the other parameter is $c$), which takes values $0<\theta <\pi $ and is
proportional to the soliton's energy (alias norm),
\begin{equation}
E\equiv \int_{-\infty }^{+\infty }\left[ |u(x)|^{2}+|v(x)|^{2}\right]
dx=8\theta \sqrt{1+c^{2}}\left( 3+c^{2}\right) ^{-1}.  \label{E}
\end{equation}
Further, $\alpha ^{-2}\equiv \frac{3}{2}+c^{2}$, $\tanh y\equiv c$, $\phi
_{0}$ is an arbitrary real constant, and
\begin{eqnarray}
X &=&\left( 1-c^{2}\right) ^{-1/2}\left( x-c\,t\right) ,\,T=\left(
1-c^{2}\right) ^{-1/2}\left( t-c\,x\right) ,  \nonumber \\
\phi (X) &=&\alpha ^{2}\sinh (2y)\mathrm{\tan }^{-1}\left\{ \tanh \left[
(\sin \,\theta )X\right] \tan \left( \theta /2\right) \right\} ,
\label{params} \\
W(X) &=&\left( \sin \,\theta \right) \,\mathrm{sech}\left[ (\sin \,\theta
)X-i\left( \theta /2\right) \right] \,.  \nonumber
\end{eqnarray}
\noindent We used these exact solutions as initial conditions to simulate
collisions between identical solitons with opposite velocities.

To consider the influence of a local defect on the collision (see Section V
below), Eqs.~(\ref{pdes}) are modified as in Refs. \cite{Weinstein} and \cite
{we}:
To consider the influence of a local defect on the collision,
Eqs.~(\ref{pdes}) are modified as in Refs. \cite{Weinstein} and
\cite{we}:
\begin{eqnarray}
iu_{t}+iu_{x}+v+\left[ \left( 1/2\right) |u|^{2}+|v|^{2}\right] u
&=&-\delta (x)\left( \Gamma u-\kappa v\right),\\
iv_{t}-iv_{x}+u+\left[ \left( 1/2\right) |v|^{2}+|u|^{2}\right] v
&=&-\delta (x)\left( \Gamma v-\kappa u\right),
\label{pdedefect}
\end{eqnarray}
where $\Gamma >0$ and $\kappa >0$ account for a local increase of the
refractive index and suppression of the Bragg reflectivity, respectively.

\section{Collisions between solitons}

\subsection{The mode of simulations}

In this section, we consider collisions between exact BG solitons (\ref
{movsol}). In a real experiment, an initially launched pulse should pass
some distance to shape itself into a soliton. As it was mentioned above, in
previously reported experiments this distance was quite small, $\sim 2$ cm
\cite{experiment}, hence this is not a big issue. Nevertheless, it is
relevant to separately simulate shaping of an initially launched
single-component pulse into a steady-shape BG soliton. This will be done
separately below in section VI.

Simulations of collisions were performed by means of the split-step
fast-Fourier-transform method. First, collisions between solitons in the
case of repulsion between them (with a phase difference $\Delta \phi
_{0}=\pi $) was considered. It was found that the solitons bounce from each
other quasi-elastically, without generation of any visible radiation or
intrinsic vibrations of the solitons, if their initial velocities $\pm c$
are small enough, and the solitons are ``light'', having a sufficiently
small value of $\theta $. Collision-induced radiation becomes quite
conspicuous if the solitons are ``heavier'' or faster, see an example in the
inset to Fig. 1. Figure 1 shows a boundary in the plane $\left( c,\theta
\right) $, above which the collision results in generation of noticeable
amount of radiation, in the case $\Delta \phi _{0}=\pi $.

Then, collisions between in-phase solitons, with $\Delta \phi _{0}=0$ (the
case of attraction), were simulated. In this case, a number of various
outcomes can be distinguished. A summary of the results is displayed in Fig.
2 in the form of a diagram in the $(c,\theta )$ plane, different outcomes
being illustrated by a set of generic examples displayed in Fig. 3.

The simplest case is the collision of solitons with small $\theta $ (region
E in Fig. 2; see also Fig. 4 below). In accordance with results reported in
Ref. \cite{Litchinitser}, these solitons collide elastically, which is
easily explained by the fact that they are virtually tantamount to NLS
solitons.

\subsection{Merger of solitons and spontaneous symmetry breaking}

The most interesting outcome of the collision is \emph{merger} of two
solitons into a single one, which takes place in the region $0\leq c<0.2$,
$0.15\pi <\theta <0.35\pi $ (area M in Fig. 2). A typical example of the
merger is shown in Fig. 3(a), its noticeable peculiarities being that the
merger takes place after multiple collisions, and the finally established
soliton demonstrates persistent internal vibrations, see the lower panel of
Fig.~3(a). As judged from the lowest panel of Fig.~3(a) [and other similar
plots], the amplitudes of these internal vibrations amount to about $10$ to
$20\%$ of the soliton amplitudes.  In this region (area M) of the values of
$\theta $, the attraction between
initially quiescent ($c=0$) in-phase solitons, which are placed at some
distance from each other, also results in their merger, see Fig. 3(b). At
the border between the regions M and E, the interaction between initially
quiescent or slow solitons results in their separation after several
collisions, which is accompanied by a conspicuous spontaneous symmetry
breaking (SSB), see an example in Fig. 3(c). Note that the SSB resembles
what was observed in a model of a dual-core fiber grating, in which the
nonlinearity and BG proper were carried by different cores \cite{Atai}. As
well as in that case, SSB may be plausibly explained by a fact that the
``lump'', which temporarily forms as a result of the attraction between the
solitons in the course of the collision between them, may be subject to
modulational instability, hence a small asymmetry in the numerical noise may
provoke conspicuous symmetry breaking in the eventual state. Indeed, it is
well known that any spatially uniform solution to Eqs. (\ref{pdes}) is
modulationally unstable \cite{MI}, and it is obvious that the instability
can extend to any sufficiently broad state.

\subsection{Quasi-elastic collisions}

Increase of $\theta $ brings one from the region M to F (Fig. 2), where
solitons collide quasi-elastically, i.e., they separate after the collision,
emerging with smaller amplitudes, see Fig. 3(d). A noticeable peculiarity of
this case is that the collision results in an \emph{increase} of the
solitons' velocities, which is seen in the change of the slope of the
contour-level plots in Fig. 3(d). We note that, pursuant to Eq. (\ref{E}),
the soliton's energy monotonically increases with $c^{2}$, therefore the
collision-induced decrease of the amplitude may be explained not only by
radiation loss, but also by the increase of the velocities. The acceleration
of the solitons due to the collision is more salient if the initial velocity
$c$ is small; for instance, initially quiescent solitons (with $c=0$)
acquire a large velocity after the interaction, see Fig. 3(e).

As for still heavier solitons, it is known that they are unstable if $\theta
>\theta _{\mathrm{cr}}\approx 1.011\cdot (\pi /2)$ \cite{Tasgal,stability}
(this value pertains to $c=0$; $\theta _{\mathrm{cr}}$ very weakly depends
on the soliton's velocity \cite{stability}). In accordance with this, in the
region D (Fig. 2) the collision triggers a strong deformation of unstable or
weakly stable solitons, see Fig.~3(f). At essentially longer times, the
strong deformation leads to destruction of the pulses.

If $\theta $ is taken in the same range as in the merger region M, i.e.,
$0.15\pi <\theta <0.35\pi $, but with a larger velocity, the collision
picture seems in an ordinary way: the solitons separate with some decrease
in their velocity, and some loss in the amplitude. If the initial velocity
is still larger, it is possible to distinguish another region, marked R in
Fig. 2, where the velocity shows no visible change after the collision, but
emission of radiation takes place.

Quasi-elastic collisions can be naturally quantified by the ratio $\theta _{
\mathrm{out}}/\theta _{\mathrm{in}}$ of the soliton's parameter after and
before the collision, and by share of the net initial energy of the solitons
which is lost (to radiation) as the result of the collision. To this end, we
performed the least-square-error fit of pulses emerging after the collision
to the exact soliton solutions (\ref{movsol}), aiming to identify the values
of $\theta _{\mathrm{out}}$, and the post-collision velocity was measured in
a straightforward way. The corresponding soliton's energy was then
calculated by means of the formula (\ref{E}).

The results of the computation are shown in Fig. 4. A noteworthy feature,
which is obvious in both panels (a) and (b), is that inelastic effects first
strengthen with the increase of $\theta _{\mathrm{in}}$ from very small
values (which correspond, as it was said above, to the NLS limit) to $\simeq
0.3\pi $, then they weaken, attaining a \emph{minimum}, which corresponds to
the most quasi-elastic collisions, at $\theta _{\mathrm{in}}\approx 0.4\pi$,
and then get stronger again, with the increase of $\theta _{\mathrm{in}}$
up to $\simeq 0.6\pi $. Past the latter value, the isolated soliton is
strongly unstable by itself, therefore detailed study of collisions becomes
irrelevant.

\section{Multiple collisions in a fiber ring}

Since the main motivation of this work is the possibility to generate a
standing pulse by dint of collisions between BG solitons, it is natural to
consider multiple collisions, that may occur between two solitons traveling
in opposite directions in a fiber loop, or if a single soliton performs
shuttle motion in a fiber-grating cavity, i.e., a piece of the fiber
confined by mirrors (in the latter case, the soliton periodically collides
with its own mirror images). An issue for experimental realization of these
schemes is to couple a soliton into the loop or cavity. Using a linear
coupler connecting the system to an external fiber may be problematic, as
repeated passage of the circulating soliton through the same coupler will
give rise to conspicuous loss. Another solution may be to add some intrinsic
gain to the system, making it similar to fiber-loop soliton lasers, where a
soliton-circulation regime may self-start \cite{laser}. It is relevant to
mention that fiber-ring soliton lasers including BG components were
investigated before \cite{BraggLaser}. Still another possibility is to use
figure-eight fiber lasers \cite{8}, in which one loop is made of BG, while
the other one provides for the gain. Detailed analysis of these schemes is,
however, beyond the scope of this paper.

We performed simulations of the multiple collisions between two identical
solitons in the loop, imposing periodic boundary conditions. Figure~5(a)
shows an example in which the multiple collisions slow down the solitons
quite efficiently, enforcing them to merge. As is seen, in this case the
solitons underwent two collisions before the merger. The initial values
$c=0.3$ and $\theta =0.3\pi $ used in this example show that the multiple
collisions in the loop help to increase the maximum initial velocity
$c_{\max }$, that admits merger of the two solitons, by a factor of $3$ (at
least) against the single-collision case, cf. Fig. 2. In fact, the largest
value of $c_{\max }$ corresponding to the multiple collisions was found to
be $\approx 0.4$, i.e., a part of the region S from Fig. 2 is absorbed into
M in the loop configuration. The evolution of the field at the central
point, $|u(x=0)|$, which is also displayed in Fig. 5(a), demonstrates that
the emerging zero-velocity pulse is again a breather, cf. Fig. 3(a).

Another example of multiple collisions in the loop is shown in Fig.~5(b),
where the solitons initially have $\theta =0.3\pi $ and $c=0.7$, belonging
to the region R of Fig. 2. In this case, the solitons hardly undergo any
slowing down due to the collisions, while they keep losing energy. Due to
the gradual decrease of $\theta $, which is related to the energy by Eq.
(\ref{E}), the solitons gradually drift to the region E (see Fig. 2), where
the collision becomes elastic.

\section{Effect of a localized defect on the collision.}

In Refs. \cite{Weinstein} and \cite{we}, it has been found that local
attractive defects can trap gap solitons. This fact suggests a possibility
that the merger of two colliding solitons might be assisted by a defect
placed at the collision point. We investigated the effect of two kinds of
local defects, which represent BG suppression or increase of the refractive
index, corresponding, respectively, to $\kappa >0$ and $\Gamma >0$ in Eqs.
(\ref{pdedefect}) (a single collision was considered in this case).

We have found that attractive defects of either type do not actually
catalyze formation of a pinned pulse that would retain a large part of the
energy of the colliding solitons. Nevertheless, a relatively small part of
the energy gets trapped by the defect, and a small-amplitude pinned soliton
appears, see an example in Fig. 6, which is displayed for the case of the
local refractive-index perturbation, i.e., $\Gamma >0$, $\kappa =0$. Local
BG suppression, accounted for by $\kappa >0$, produces a similar effect. We
have also checked that repulsive local defects (negative $\Gamma $ or
$\kappa $) do not produce any noticeable effect either.

\section{Special effects in the single-soliton dynamics}

\subsection{Transformation of an input pulse into a Bragg-grating soliton}

As it was mentioned above, signals which are coupled into a fiber grating in
a real experiment are not ``ready-made'' BG solitons, but rather pulses of a
different form, which should shape themselves into solitons. After that, one
can consider collisions between them, as it was done above. For this reason,
it makes sense to specially consider self-trapping of BG solitons from a
standard input pulse in the form of the NLS soliton,
\begin{equation}
u_{0}(x)=\eta \,\mathrm{sech}(\eta x)\exp (i\kappa x),\,v_{0}(x)=0.
\label{NLS}
\end{equation}
where $\eta $ and $\kappa $ are two arbitrary real parameters. The energy of
the pulse (\ref{NLS}), defined as per Eq. (\ref{E}), is $E_{0}=2\eta $.

Transformation of the pulse into a BG soliton was simulated directly within
the framework of Eqs. (\ref{pdes}). The results are summarized in Fig. 7, in
the form of plots showing the share of the initial energy lost into
radiation, cf. Fig. 4(b). A noteworthy feature revealed by the systematic
simulations is that, with the increase of the parameter $\eta $, that
measures the amplitude and inverse width of the initial pulse (\ref{NLS}),
the energy-loss share first decreases, attaining an absolute minimum at
$\eta \simeq 0.8-1.0$, and then quickly increases. The fact that the relative
energy loss becomes very large for large $\eta $ is easy to understand, as
the initial energy of the pulse (\ref{NLS}) increases indefinitely with
$\eta $, while the energy of the emerging stable BG pulse, with $\theta \leq
1.011\cdot (\pi /2)$ and $c^{2}<1$, cannot exceed (in the present notation) $E_{\max }=
\sqrt{2}\pi $, see Eq. (\ref{E}) [we did not observe formation of more than
one BG soliton from the initial pulse (\ref{NLS})]. Thus, an optimum shape
of the sech input signal, which provides for the most efficient generation
of the BG soliton, is suggested by these results.

\subsection{Decay of the soliton in a finite-length fiber grating with free
ends}

In any experiment (unless the fiber loop or cavity are used), a standing
soliton will be created in a fiber grating with open edges. Then, some
energy leakage will take place through free ends of the fiber segments. From
the exact solution (\ref{movsol}) it follows that the leakage is
exponentially small if the segment's length $l$ is much larger than the
soliton's spatial width, which is $\sim 1$ mm in a typical situation \cite
{experiment,we2}. Moreover, the energy leakage through loose end can be
easily compensated (along with intrinsic fiber loss) by local gain
\cite{we2}. Nevertheless, it is an issue of interest to
find the soliton's decay rate due to the leakage.

We addressed the issue, simulating Eqs. (\ref{pdes}) with the free boundary
conditions, $u_{x}=v_{x}=0$, set at the edges of the integration domain. In
Fig. 8, we show the decay of the soliton's amplitude in time, for different
values of the domain's length, with initial $\theta_{in} =0.51$.  The initial
increase of the amplitude is a
result of temporary self-compression of the pulse due to its interaction
with the edges. As a reference, we mention that, in the case of the shortest
fiber grating considered here, with $l=8$, it takes the time $t=42.2$ for
the decrease of the amplitude by a factor of $e$.

\section{Conclusion}

We have presented results of systematic studies of collisions between moving
solitons in fiber gratings. Various outcomes of the collision were
identified, the most interesting one being merger of the solitons into a
single zero-velocity pulse, which suggests a way to create pulses of
``standing light''. The merger occurs for solitons whose energy takes values
between $0.15$ and $0.35$ of its maximum value, while the velocity is
limited by $c_{\max }\approx 0.2$ of the limit velocity. If the energy is
larger, another noteworthy outcome is acceleration of the solitons as a
result of the collision, especially when their initial velocities are small.
In the case when the solitons pass through each other, inelasticity of the
collision was quantified by the relative energy loss. If the energy exceeds
the soliton's instability threshold, the collision results in strong
deformation of the solitons, which is followed by their destruction.
Simulations of multiple collisions between two solitons in the fiber-loop
configuration show that the largest initial velocity admitting the merger
increases to $c\leq c_{\max }\approx 0.4$ of the limit velocity. It was also
shown that attractive local defects do not alter the overall picture,
although a small-amplitude trapped pulse appears in this case. Finally,
specific effects were investigated in one-soliton dynamics, such as
transformation of a single-component pulse into a Bragg-grating soliton, and
decay of the soliton in a finite-length fiber grating due to the energy
leakage through loose edges.

\section*{Acknowledgement}

One of the authors (B.A.M.) appreciates hospitality of the Optoelectronic
Research Centre at the Department of Electronic Engineering, City University
of Hong Kong.

\newpage

\section*{Figures}

\begin{figure}[hb]
\includegraphics[scale=0.7]{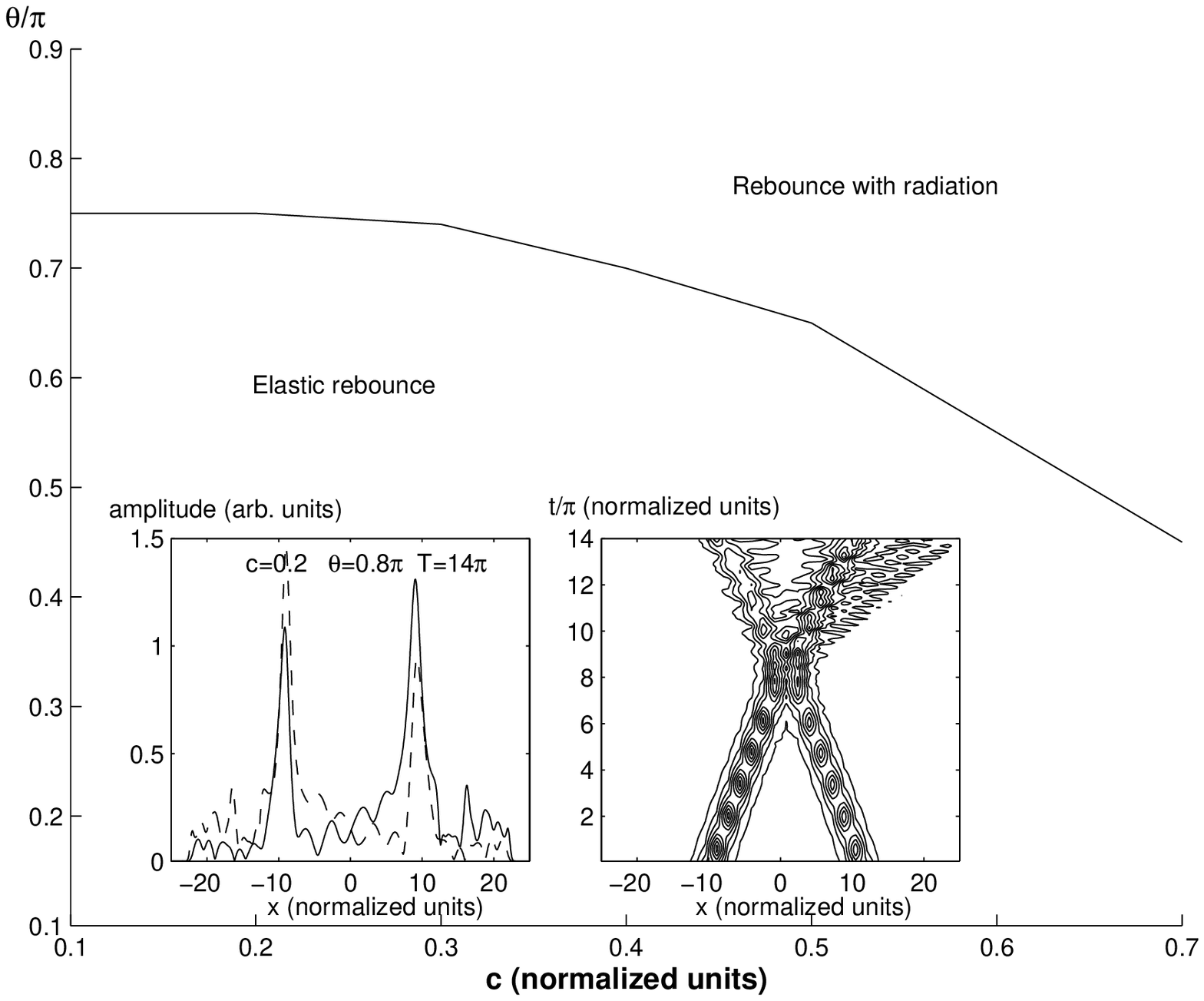}
\caption{The border separating regions in the plane ($c, \protect\theta _{
\mathrm{in}}$) where, in the case of repulsion, the collision is elastic, or
generates significant radiation loss. An example of the collision of the
latter type is given in the inset, in which the left and right panels show,
respectively, the waveforms $|u(x)|$ and $|v(x)|$ (solid and dashed lines)
at the end of the simulation ($t=14\protect\pi$), and the evolution of the
field $|u(x,t)|$ in terms of level contours. Intrinsic ``oscillations"
of the solitons before the collision in the inset is an artifact due to
mismatch between plotting sampling and the numerical grid used for
the simulations.}
\end{figure}

\begin{figure}[hb]
\includegraphics[scale=0.7]{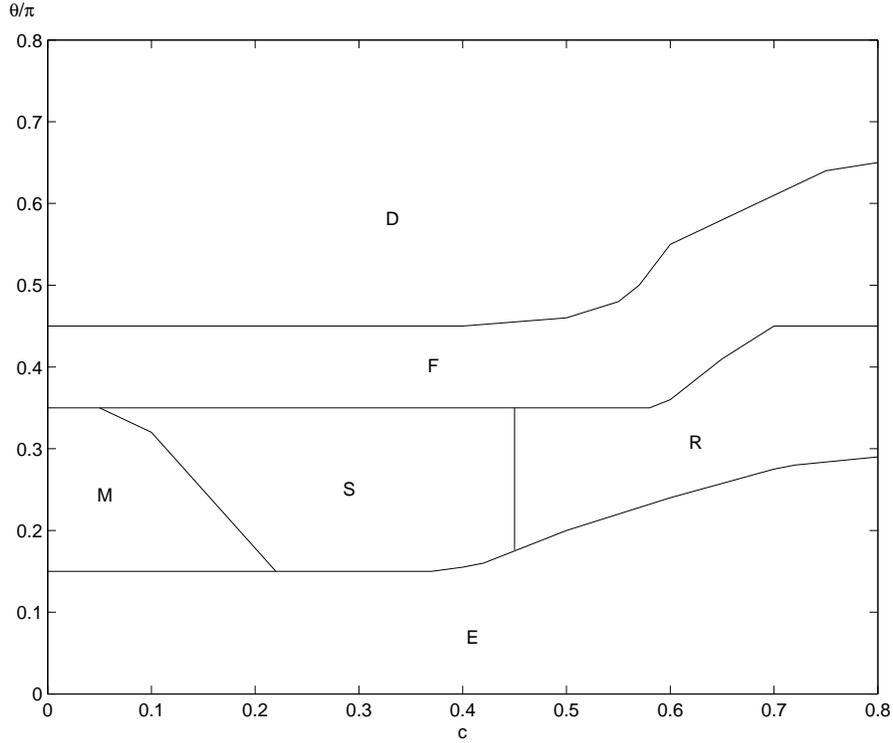}
\caption{ A diagram in the plane ($c, \protect\theta _{\mathrm{in}}$) for
different outcomes of the collision between in-phase solitons. In the region
E the collision is elastic. In the region M, the solitons merge into a
single pulse. In the region S, they separate with velocities smaller than
they had before the collision. In the region R, the velocities are not
affected by the collision, but conspicuous radiation losses are observed. In
the region F, large radiation loss takes place, and the velocities increase
after the collision. In the region D, the collision leads to strong
deformation of the solitons.}
\end{figure}

\newpage

\begin{figure}[ht]
\centerline{\scalebox{0.7}{\includegraphics{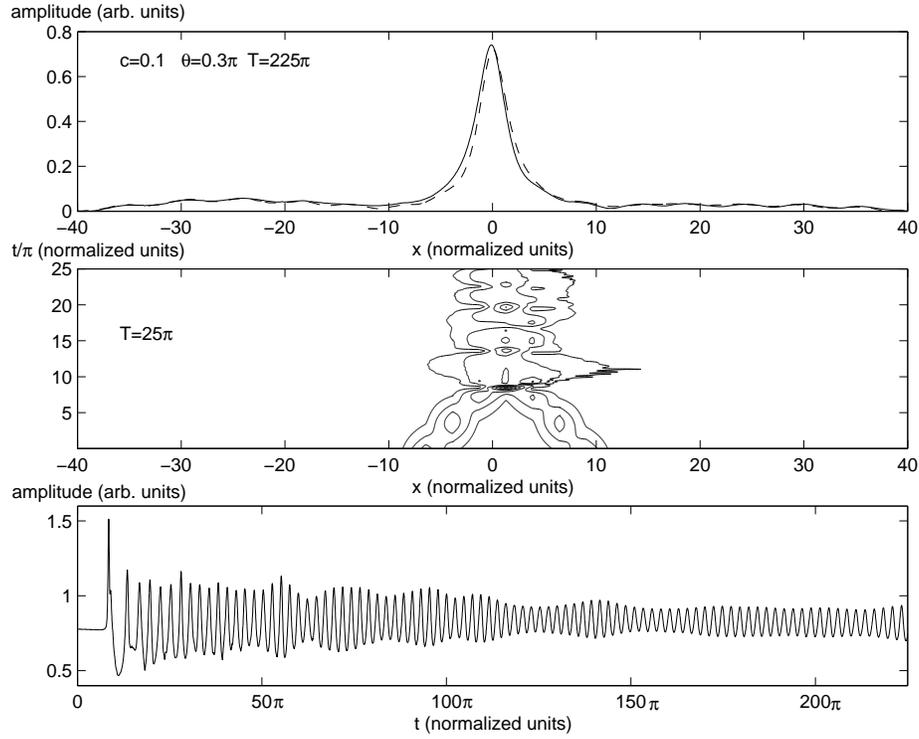}}}
\caption{Typical examples of the collision between in-phase solitons. (a)
Merger of the solitons in the region M in Fig. 3. They collide several times
before the merger, which is accompanied by emission of radiation. The lower
panel exhibits persistent vibrations of the field amplitude $|u(x=0,t)|$.
Here and below, the middle and upper panels show, respectively, the
evolution at a relatively early stage ($t=25\protect\pi$), and the single
pulse emerging at $t=225\protect\pi$.}
\end{figure}

\addtocounter{figure}{-1}

\begin{figure}[ht]
\centerline{\scalebox{0.7}{\includegraphics{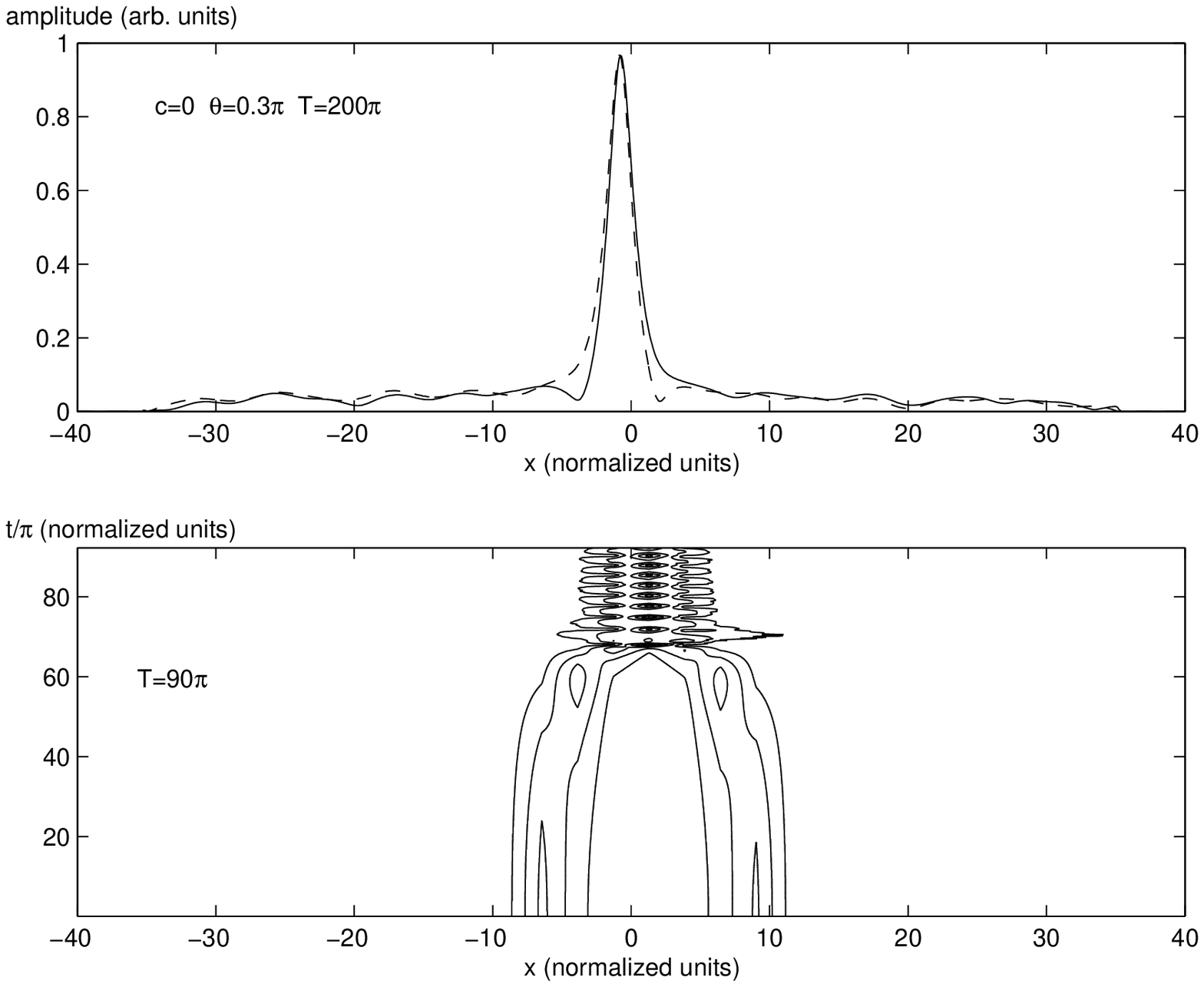}}}
\caption{(b) Merger of initially quiescent solitons ($c=0$). The lower and
upper panels show the evolution at $t<90\protect\pi$, and the emerging
single pulse at $t=200\protect\pi$.}
\end{figure}

\addtocounter{figure}{-1}

\begin{figure}[hb]
\centerline{\scalebox{0.7}{\includegraphics{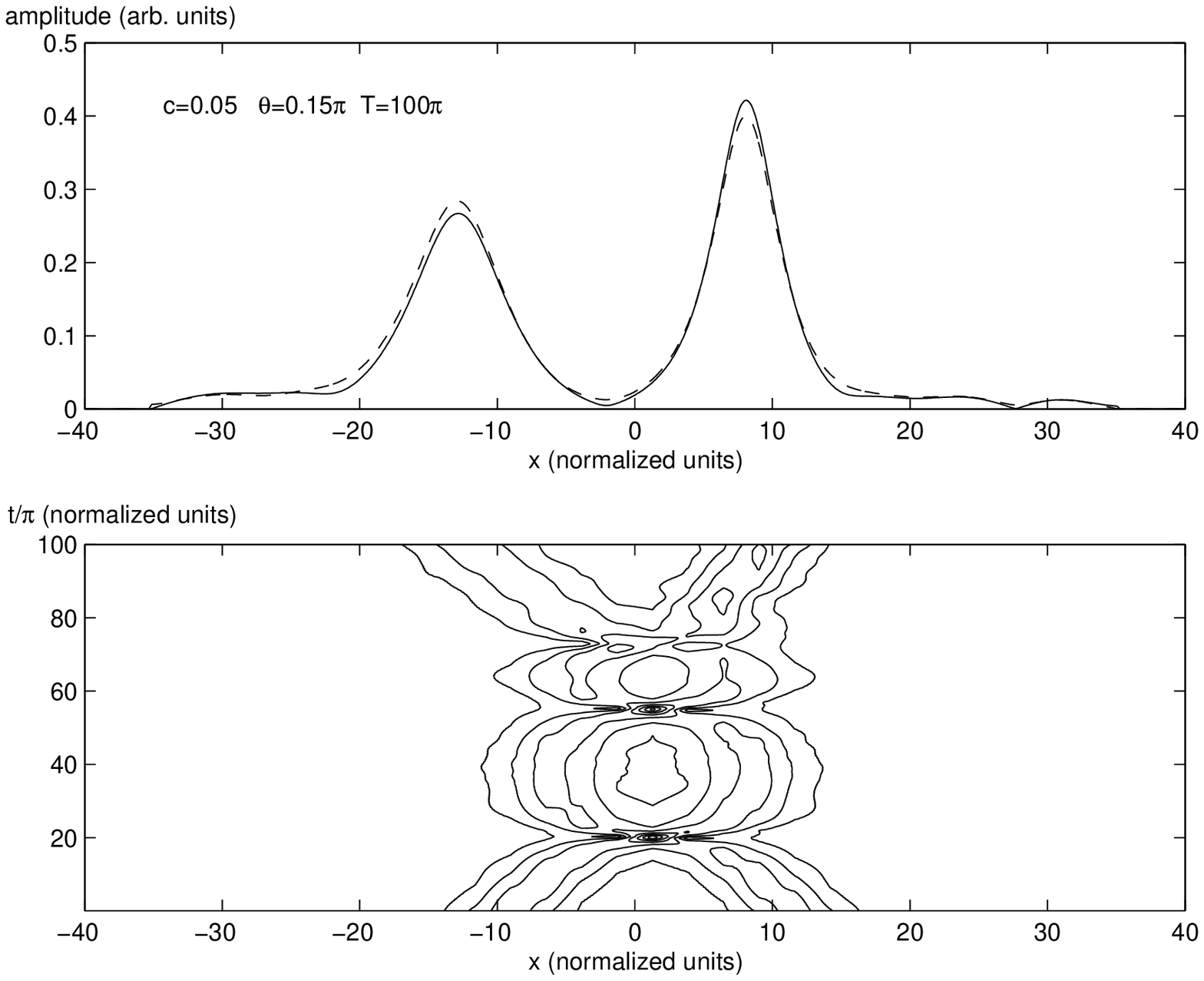}}}
\caption{(c) At the lower edge of region M (Fig.~3), solitons undergo
multiple collisions before they finally separate. Spontaneous symmetry
breaking is evident in the final state.}
\end{figure}

\addtocounter{figure}{-1}

\begin{figure}[hb]
\centerline{\scalebox{0.7}{\includegraphics{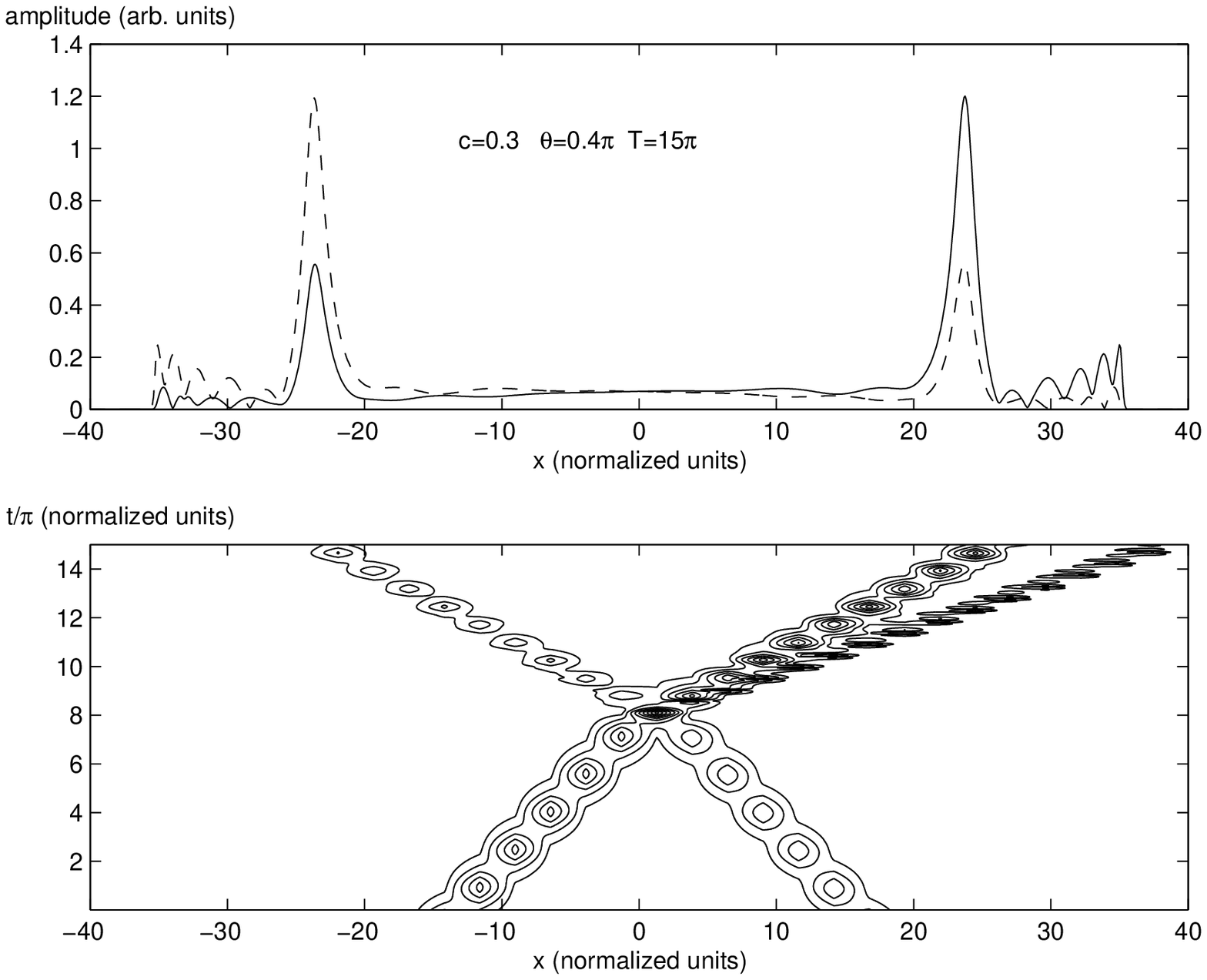}}}
\caption{(d) Collision between relatively heavy solitons leads to emission
of radiation jets and increase of the velocities (region F in Fig. 3).}
\end{figure}

\addtocounter{figure}{-1}

\begin{figure}[hb]
\centerline{\scalebox{0.7}{\includegraphics{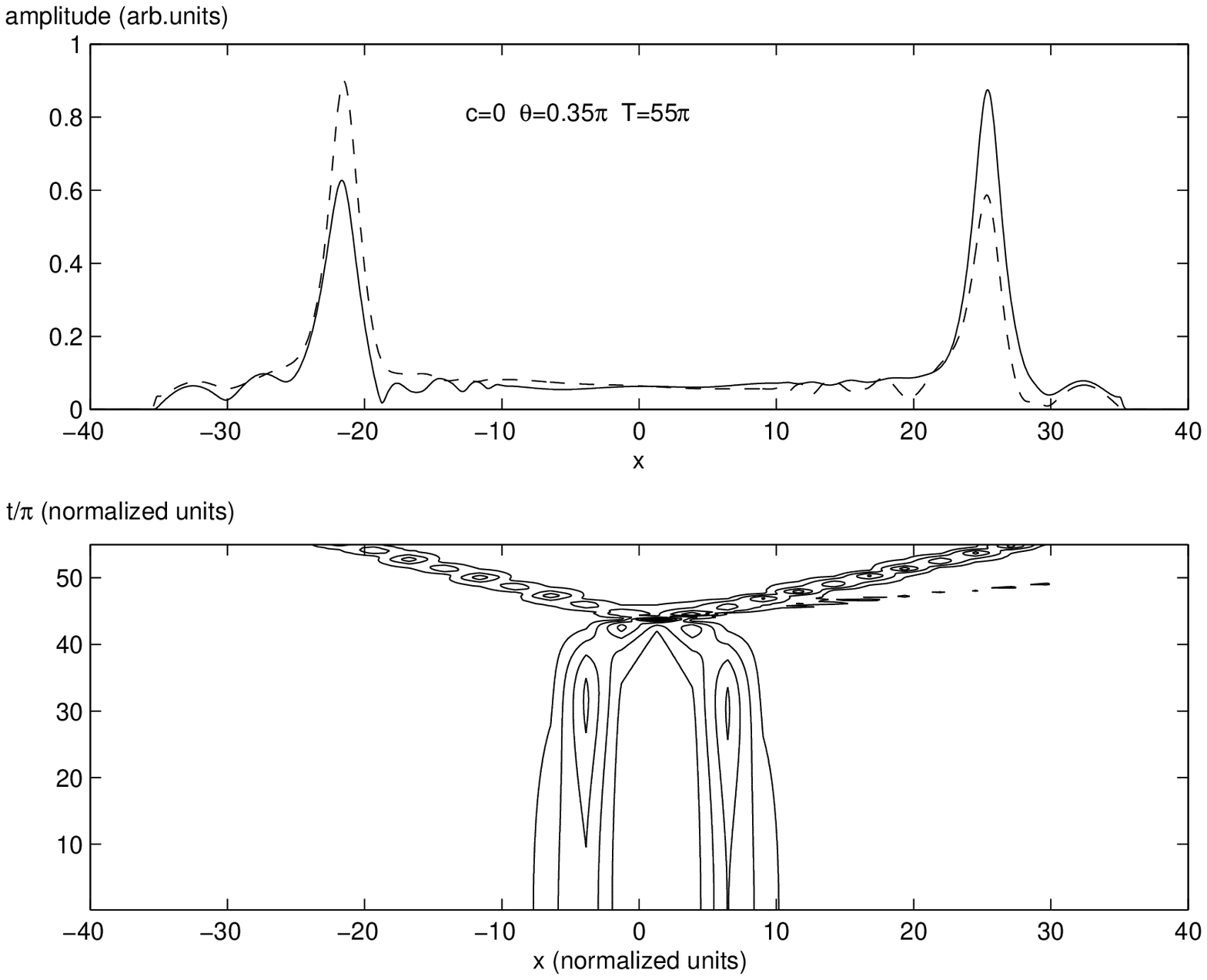}}}
\caption{(e) Interaction between two initially quiescent solitons in the
region F (Fig.~2).}
\end{figure}

\addtocounter{figure}{-1}

\begin{figure}[hb]
\centerline{\scalebox{0.7}{\includegraphics{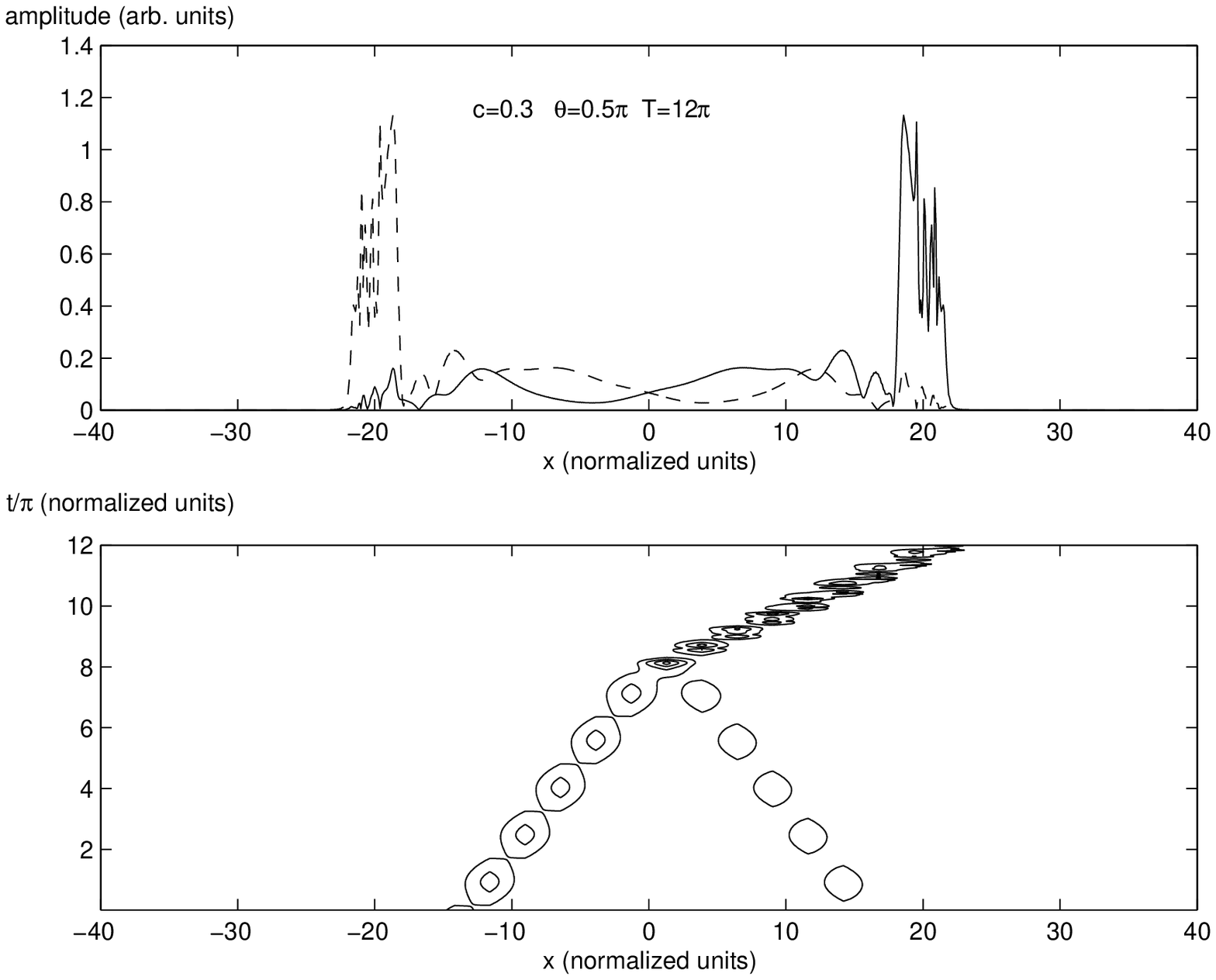}}}
\caption{(f) Collision between heavy solitons which are weakly stable or
unstable (region D in Fig.~2) results in strong deformation of the pulses,
which is followed by their destruction (not shown here).}
\end{figure}

\begin{figure}[hb]
\centerline{\scalebox{0.7}{\includegraphics{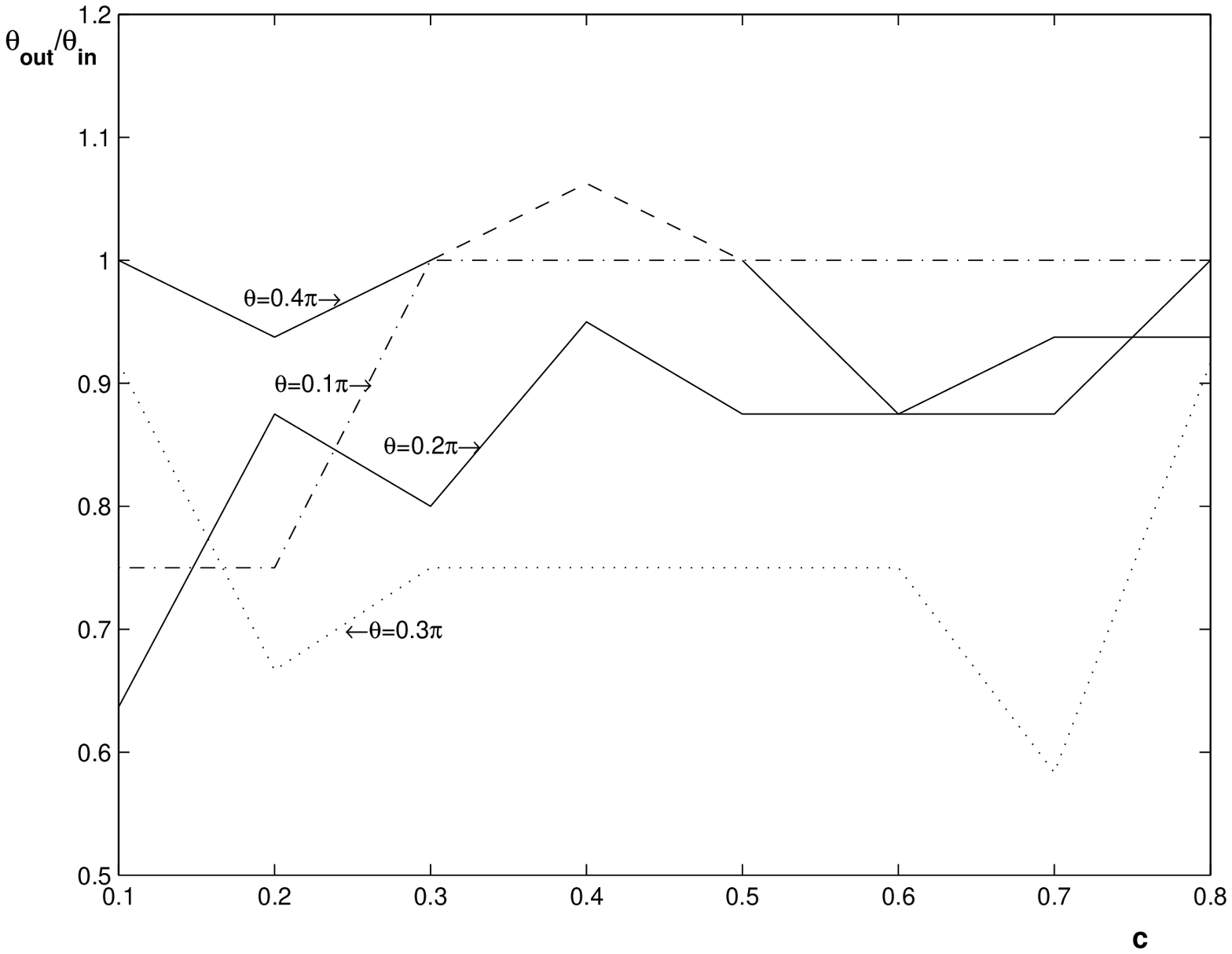}}}
\caption{(a) The ratio of the post-collision soliton's parameter
$\theta_{\mathrm{out}}$, found from the least-square-error fit of the
emerging pulse to the analytical waveforms (\ref{movsol}), to the initial
value $\protect\theta_{\mathrm{in}}$. In this and next panels, the ratio is
shown vs. the initial velocity $c$ at different fixed values of
$\theta_{\mathrm{in}}$. The portion of the line corresponding to
$\theta_{\mathrm{in}}=0.4\protect\pi$ with $\protect\theta_{\mathrm{out}}/
\protect\theta_{\mathrm{in}}>1$, which formally contradicts the energy
conservation, is explained by the fact that in this case the actual shape of
the emerging pulse is not very close to the analytical one, being more
narrow.}
\end{figure}

\addtocounter{figure}{-1}

\begin{figure}[hb]
\centerline{\scalebox{0.7}{\includegraphics{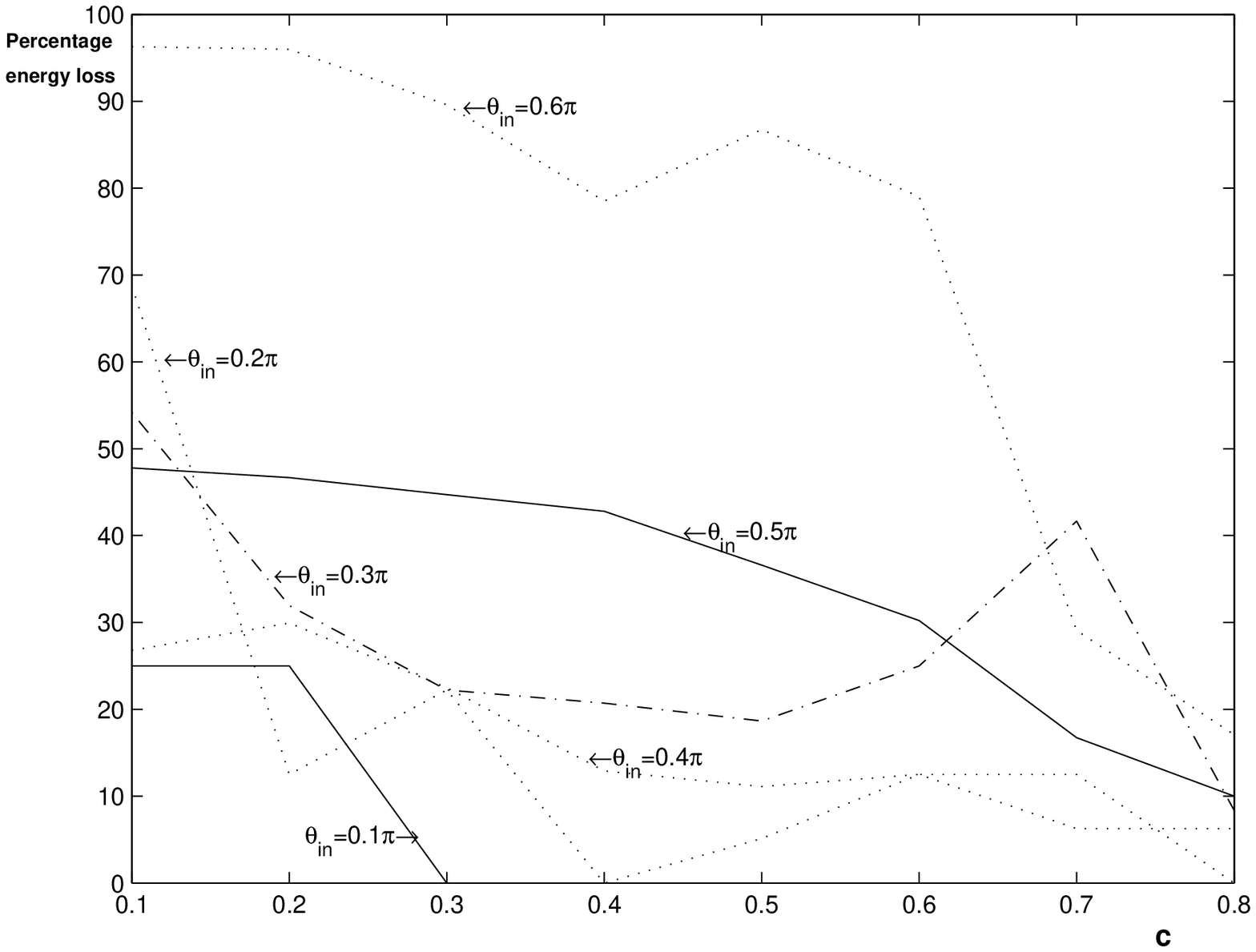}}}
\caption{(b) The relative energy loss due to the collision of two solitons.}
\end{figure}

\begin{figure}[hb]
\centerline{\scalebox{0.7}{\includegraphics{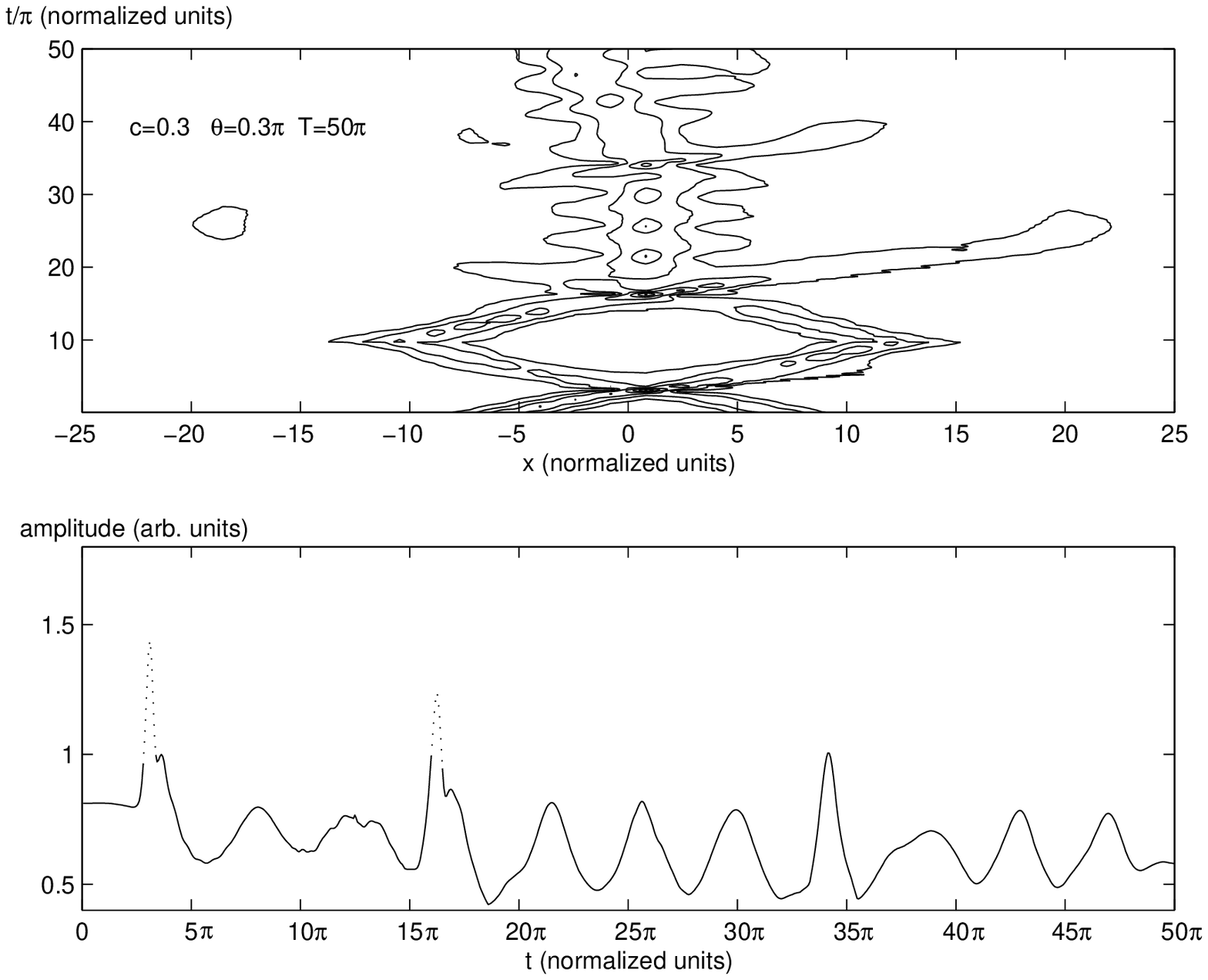}}}
\caption{(a) Multiple collisions between two solitons with the initial value
$\protect\theta=0.3\protect\pi$, and initial velocity $\pm 0.3$ in the loop
configuration. The upper and lower panels, respectively, show the global
evolution of the field $|u(x,t)|$, and the evolution of its maximum. In the
lower panel, the dotted parts of the curve mark two collisions (maximum
overlappings) between the two solitons before they merge into a single
pulse. }
\end{figure}

\addtocounter{figure}{-1}

\begin{figure}[hb]
\centerline{\scalebox{0.7}{\includegraphics{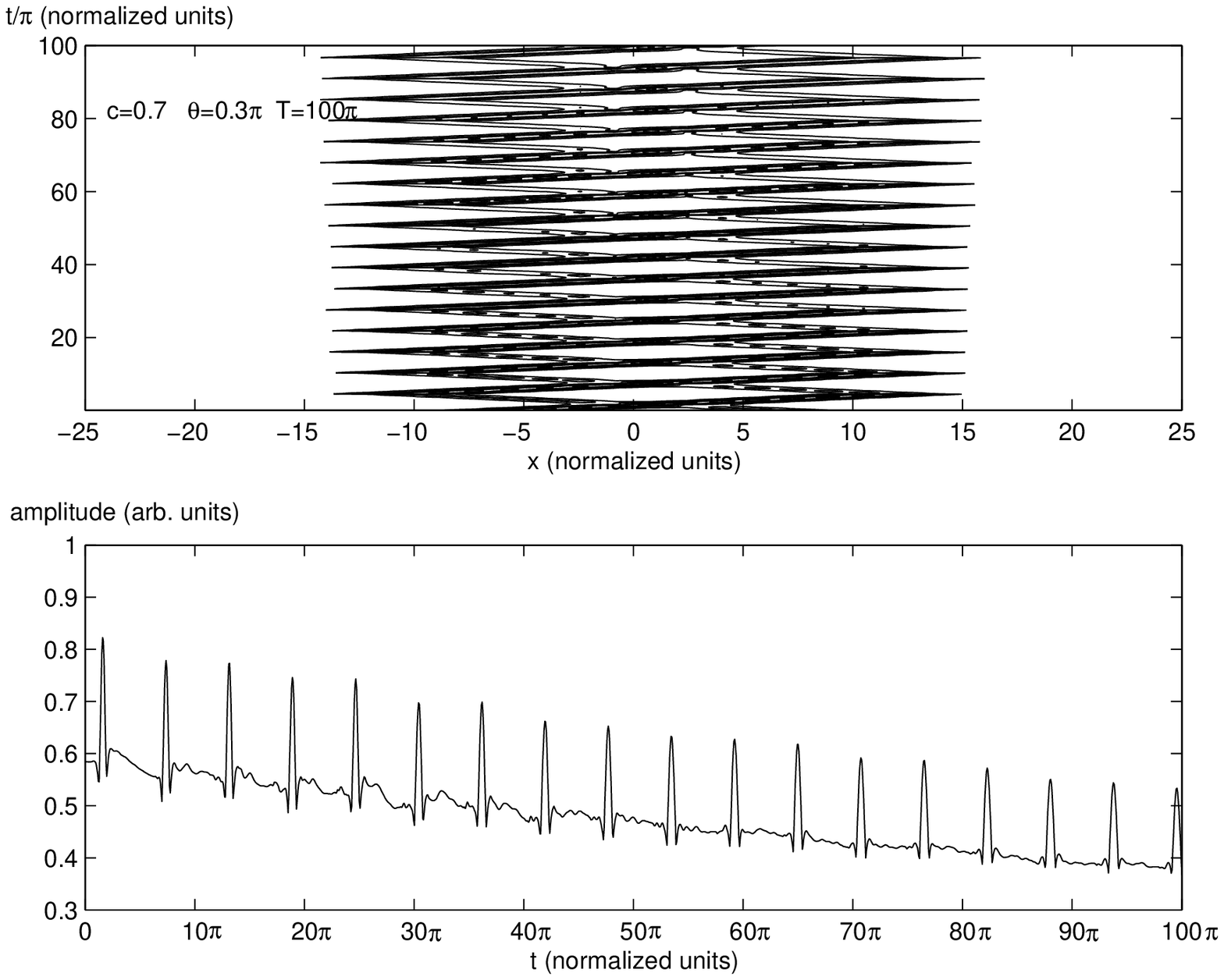}}}
\caption{(b) Multiple collisions between solitons with the initial value
$\protect\theta=0.3\protect\pi$ and initial velocities $\pm 0.7$ in the loop
configuration.}
\end{figure}

\begin{figure}[hb]
\centerline{\scalebox{0.7}{\includegraphics{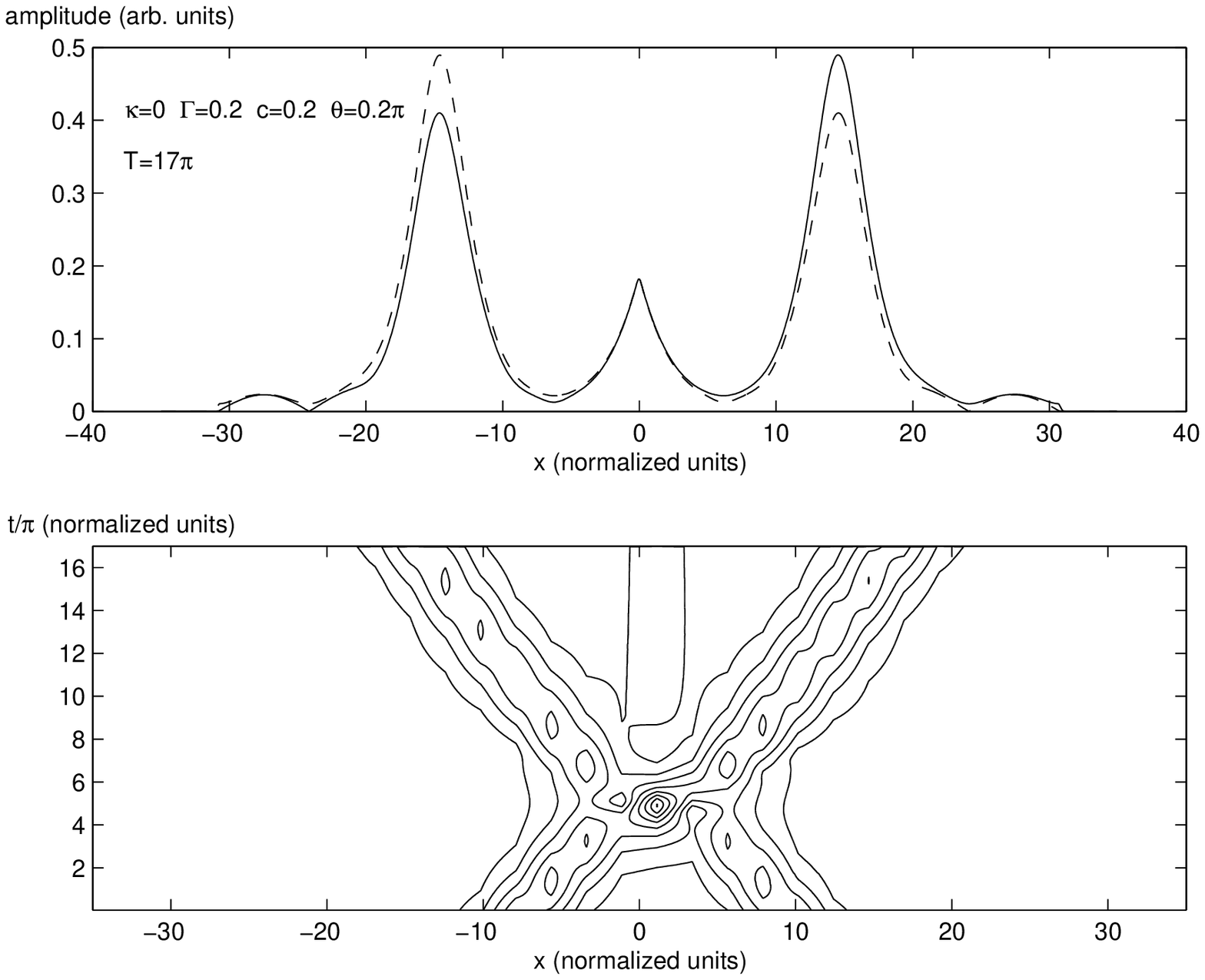}}}
\caption{The collision between solitons with $\protect\theta=0.2\protect\pi$
and velocities $c=\pm 0.2$ in the case when a local perturbation of the
refractive index, with $\Gamma=0.2$ [see Eqs. \ref{pdedefect})], is placed
at the collision point. The defect traps a small-amplitude soliton.}
\end{figure}

\begin{figure}[hb]
\centerline{\scalebox{0.7}{\includegraphics{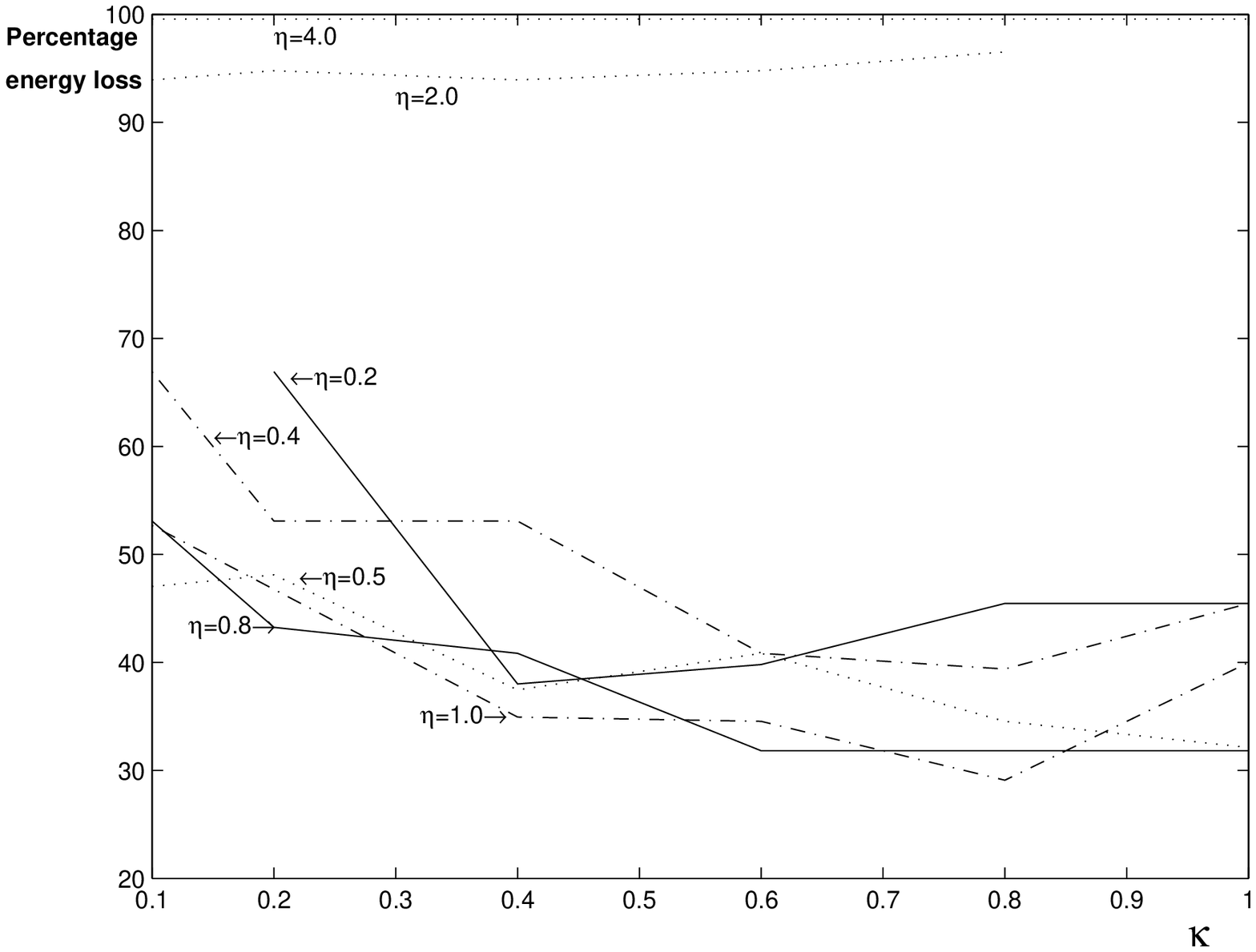}}}
\caption{The relative energy loss in the process of self-trapping of the
Bragg-grating soliton from the initial pulse (\ref{NLS}) vs. the parameter
$\protect\kappa$ at fixed values of the amplitude $\protect\eta$.}
\end{figure}

\begin{figure}[hb]
\centerline{\scalebox{0.7}{\includegraphics{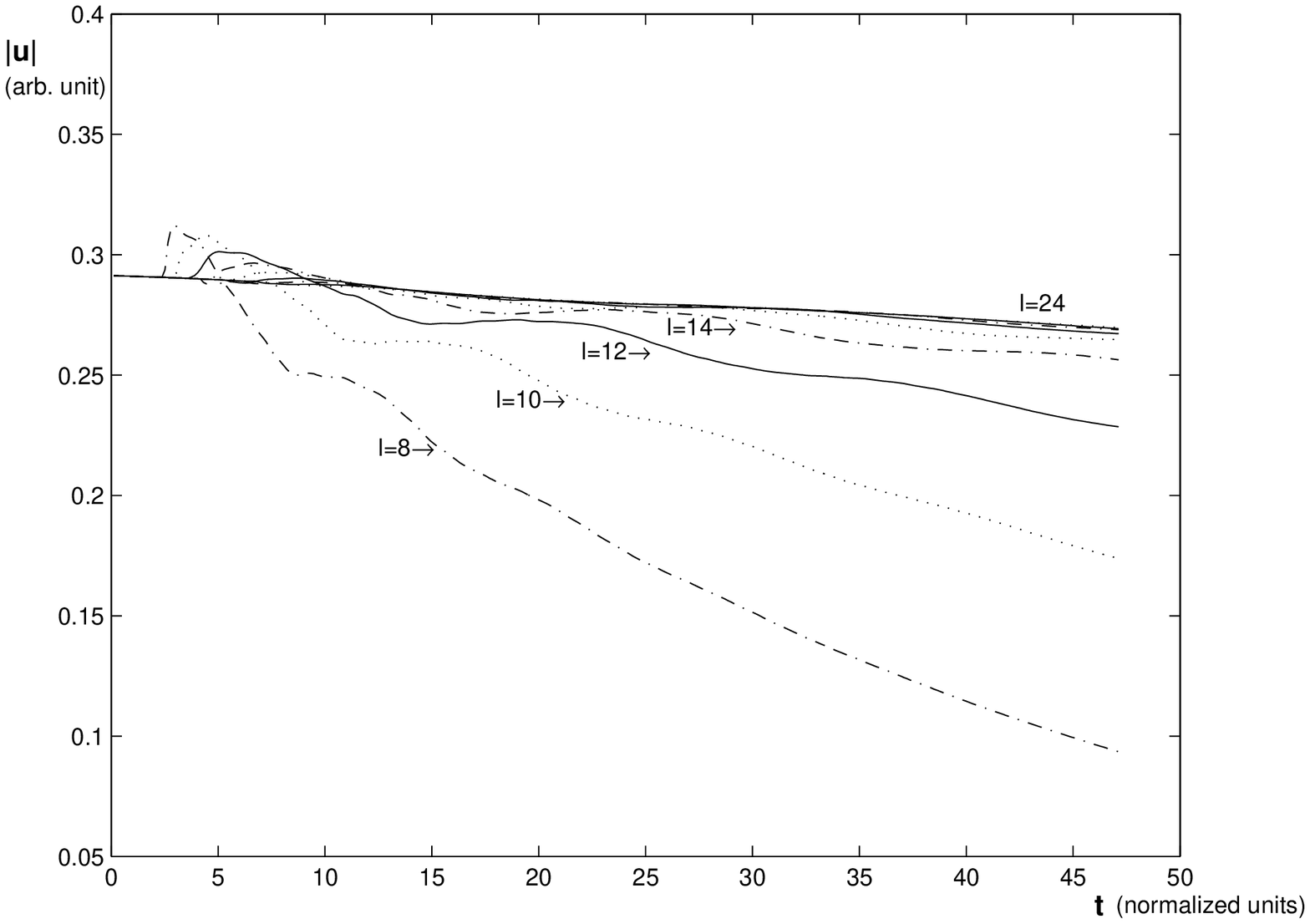}}}
\caption{Decay of the field $|u(x,t)|$ at the central point of the finite
fiber grating of length $l$ due to the energy loss through free ends.}
\end{figure}

\end{document}